# Creating a zero-order resonator using an optical surface transformation

Fei Sun [1, 2], Xiaochen Ge [1], and Sailing He [1, 2, *]

1 State Key Laboratory of Modern Optical Instrumentations, Centre for Optical and Electromagnetic Research, JORCEP, East Building #5, Zijingang Campus, Zhejiang University, Hangzhou 310058, China

2 Department of Electromagnetic Engineering, School of Electrical Engineering, Royal Institute of Technology (KTH), S-100 44 Stockholm, Sweden

* Correspondence and requests for materials should be addressed to S. He (sailing@kth.se ).

**Abstract**

A novel zero-order resonator has been designed by an optical surface transformation (OST) method. The resonator proposed here has many novel features. Firstly, the mode volume can be very small (e.g. in the subwavelength scale). Secondly, the resonator is open (no reflecting walls are utilized) and resonant effects can be found in a continuous spectrum (i.e. a continuum of eigenmodes). Thirdly, we only need one homogenous medium to realize the proposed resonator. The shape of the resonator can be a ring structure of arbitrary shape. In addition to the natural applications (e.g. optical storage) of an optical resonator, we also suggest some other applications of our novel optical open resonator (e.g. power combination, squeezing electromagnetic energy in the free space).

**Introduction**

Cavities/resonators have been widely utilized to confine electromagnetic energy [1, 2]. The mode volume and the quality factor Q are two important properties of an optical cavity/ resonator, reflecting its ability to confine the light in the spatial domain and the time domain, respectively. Surface plasmon polariton cavities can confine the light in a subwavelength scale [3]. However, the quality factor Q is limited due to the loss in the metal. Dielectric cavities can achieve a high Q [4], although the whole structure (e.g. a photonic crystal cavity) is usually much larger than the operating wavelength. One approach for compromise is to combine the metals and dielectric materials together to form a composite cavity, similarly to designing a hybrid dielectric-plasmonic waveguide that can confine the light on a subwavelength scale with low loss (see e.g. [5]).

Transformation optics (TO) is a powerful theoretical tool that can be utilized to design many novel optical/electromagnetic devices with pre-designed functions [6-8]. V. Ginis's group has used TO to design many novel optical cavities that can confine the light on a subwavelength scale with an extremely high Q factor [9, 10]. Many different types of coordinate transformations have been given in their studies, and the theory on how to design an optical cavity/resonator with an unlimited quality factor and subwavelength mode volume by the coordinate transformations is available. There are also some other studies on designing an optical cavity by TO [11, 12]. These cavities designed by TO are open cavities, which can confine the light wave inside without any reflecting walls. For the traditional cavity/resonator (e.g. with some reflecting walls), a standing wave is formed inside the cavity/resonator at the resonance frequency. For an open cavity/resonator, the light path is cancelled inside the cavity. We should also note that in addition to using TO, open cavities can also be designed by using alternating positive and negative refraction index media [13, 14].

In this paper we use optical surface transformation (OST), a novel method for designing optical devices [15], to design a novel zero-order optical resonator that has the following features. Firstly, the size of the resonator can be either very large or very small compared to the working wavelength (i.e. a subwavelength mode volume can be achieved). Secondly, the resonant frequencies can be a continuous spectrum in theory. Note that in practice the resonant frequencies will be influenced by the dispersion properties of the real materials used (e.g. metamaterials or photonic crystals). Thirdly, the proposed resonator is an open resonator, which does not contain any reflecting walls. The resonant effect is due to the cancellation of the optical path like in other open cavities [14].

Compared with the optical cavities/resonators designed by TO [9-12], the optical resonator designed by OST here have many other special features. Firstly, we do not need any detailed coordinate transformation. The designing process with an OST is very simple: we just need to choose the shape of the resonator, and the medium inside the resonator will be naturally determined, as we will explain in detail later. Secondly, the shape of our resonator can be an arbitrarily shaped ring structure (not necessary a circular ring structure). We can choose resonators of different shape according to different forms of excitation. Thirdly, we only need one homogeneous anisotropic medium without any left-hand materials to realize the open resonator proposed in this paper. The optic-null medium (ONM), which has been experimentally realized in microwave frequencies [16, 17], is such a medium.

**Method**

The optical surface transformation (OST) has been recently proposed in a short paper [15]. Here we just use the final conclusion given in [15] to design the novel zero-order optical resonator. As shown in [15], two arbitrarily shaped surfaces connected by the optic-null medium (ONM) can perform equivalently (i.e. they correspond to the same surface in the reference space, and the wave propagates from one surface to the other without any phase delay). An ONM is a highly anisotropic medium whose relative permittivity and permeability are extremely large along its main axis and nearly zero in other orthogonal directions. Such an ONM has been experimentally demonstrated by metamaterials [16, 17]. For example, the ONM with the main axis in the $x$ direction can be expressed by: $\varepsilon_x = \mu_x = 1/\Delta$, $\varepsilon_y = \mu_y = \varepsilon_z = \mu_z = \Delta$, $\Delta \to 0$. The ONM with the main axis in the radial direction can be given by: $\varepsilon_r = \mu_r = 1/\Delta$, $\varepsilon_\theta = \mu_\theta = \varepsilon_z = \mu_z = \Delta$, $\Delta \to 0$.

We first show that two arbitrarily shaped surfaces linked by the ONM enclosed by two smooth curves also perform equivalently (see Figure 1(a)) and then conclude that if the two curves are conformal (i.e. the tangential directions of the two conformal smooth curves should be the same at each part), the main axis of the ONM inside the two conformal smooth curves is the same as the tangential direction of the curves. This can be proved by dividing the whole ONM into many small regions along the normal direction of one smooth curve $\Gamma_1$ (see Fig. 1 (a)). We can set up a local Cartesian coordinate system in each small region (the local $x$ coordinate variable is along the tangential direction of one curve $\Gamma_1$, and the local $y$ coordinate variable is along the normal direction). In each small region, two surface elements (e.g. $\Delta S_i$ and $\Delta S_{i+1}$, or $\Delta S_j$ and $\Delta S_{j+1}$), linked by the ONM with the main axis along the local $x$ direction, perform equivalently. This relationship will transfer from $S_1$ to $S_2$, and hence surfaces $S_1$ and $S_2$ perform equivalently (i.e. any point source on surface $S_1$ will produce a corresponding image on surface $S_2$, and the light wave propagating from $S_1$ to $S_2$ will not produce any phase delay). In each small region, the coordinate transformation between the reference space and the real space can be given as (see Figure 1(b)):

$$\begin{cases} x = \dfrac{d}{\Delta} x_0 \\ y = (\dfrac{M-1}{\Delta} x_0 + 1) y_0, \\ z = z_0 \end{cases} \quad (1)$$

where $(x, y, z)$ and $(x_0, y_0, z_0)$ denote the coordinates in the real and reference space, respectively, and $M$ determines the compression factor along the $y$ direction. The relative permittivity and permeability in this small region can be calculated with the help of TO:

$$\varepsilon = \mu = \begin{bmatrix} \dfrac{d}{\Delta Q} & \dfrac{P}{Q} & 0 \\ \dfrac{P}{Q} & \dfrac{\Delta}{d} \dfrac{P^2 + Q^2}{Q} & 0 \\ 0 & 0 & \dfrac{\Delta}{d} \dfrac{1}{Q} \end{bmatrix}, \quad (2)$$

where $P = \dfrac{M-1}{Q} y, Q = \dfrac{M-1}{d} x + 1$. When $\Delta \to 0$, the medium will reduce to the ONM. However if $M \neq 1$ (i.e., there is some scaling along the normal direction of two conformal curves, which means that the two curves are not conformal), the main axis of the ONM described by Eq. (2) will not be along the local $x$ direction (i.e. the tangential direction of curve $\Gamma_1$). If we assume the tangential directions of the two curves (i.e. $\Gamma_1$ and $\Gamma_2$) are the same in each small region, which also means that the area of the cross section between two curves $\Gamma_1$ and $\Gamma_2$ is a constant in our design (i.e. $M=1$ in all small regions), then Eq. (2) reduces to

$$\varepsilon = \mu = diag(\dfrac{d}{\Delta}, \dfrac{\Delta}{d}, \dfrac{\Delta}{d}) \xrightarrow{\Delta \to 0} diag(\infty, 0, 0). \quad (3)$$

Eq. (3) shows that the medium in each small region (e.g. region between $\Delta S_i$ and $\Delta S_{i+1}$ in Figure 1(a)) is always the ONM with main axis along the local $x$ direction (i.e.

the tangential direction of the two conformal curves) if two curves (i.e. $\Gamma_1$ and $\Gamma_2$) are conformal (i.e. the tangential directions of the two curves are the same in each small region).

Consider one special case in which the shape of the two surfaces $S_1$ and $S_2$ are the same, and imagine that the ONM connected to the two arbitrarily shaped surfaces performs like a cable that can be bent smoothly. If we connect $S_1$ and $S_2$ together, we will obtain a closed loop filled with the ONM (see Figure 1(c)), which will perform like a novel optical resonator. The shape of such a loop can be arbitrary (provided that the inner and outer boundaries of this loop are conformal smooth curves).

First we introduce a simple way to design an open resonator by OST. We begin by choosing the shape of a closed loop whose inner and outer boundaries are conformal smooth curves (i.e., the two curves always have the same local tangential direction). The medium inside the loop is the ONM whose main axis is along the tangential direction of the two conformal curves. Such a closed loop filled by the ONM performs like a special optical open resonator. A circular ring structure is the simplest shape.

The way to design an open resonator by the OST is not limited to the above simple way. Actually the OST can make two arbitrarily shaped surfaces $S_1$ and $S_2$ equivalent (see Figure. 1(d)). We can also design some ONM to make equivalent another pair of two surface $S_1$' and $S_2$', which have exactly the same as shapes as $S_1$ and $S_2$, respectively. If we connect $S_1$ and $S_1$', $S_2$ and $S_2$' together, respectively, we will obtain a closed loop filled with ONM, performing like an open resonator (see Figure. 1(d)). $\Gamma_1$ and $\Gamma_2$ are not necessarily smooth conformal curves (i.e. they can have some sharp corners where we cannot define the tangential direction). An example designed by this method is shown in Figure 4(b).

**Explicit solution for a special case of concentric ring resonator**

We use the finite element method (FEM) to simulate the performance of the proposed resonator. For simplicity, we consider a 2D concentric ring resonator (i.e. the inner and the outer boundaries of the cavity are both circles, the main axis of the ONM filled in the cavity is in the $\theta$ direction). The structure of this open resonator is given in Figure 2(a). If we set a unit line current at the center of the concentric ring, the mode in the 2D concentric ring resonator is excited (see Figure 2(b)). We should note that there is no resonant frequency for this resonator (i.e. if we change the frequency of the line current, the mode in the open resonator can still be excited). We can also analyze the eigenmode of the resonator and explain this phenomenon analytically due to the simple geometry we choose here. The Helmholtz equation for a 2D TE wave propagation in an anisotropic medium in a cylindrical coordinate system can be given by:

$$\frac{1}{r}\frac{\partial}{\partial r}\left(\frac{r}{\mu_\theta}\frac{\partial}{\partial r}\right)E_z + \frac{1}{r^2}\frac{\partial}{\partial \theta}\left(\frac{1}{\mu_r}\frac{\partial}{\partial \theta}\right)E_z + k_0^2\varepsilon_z E_z = 0, \qquad (4)$$

where $\mu_r$ and $\mu_\theta$ are the relative permeabilities in the radial and tangential directions, respectively, and $\varepsilon_z$ is the relative permittivity in the $z$ direction. The ONM is a homogenous medium, and hence we can rewrite Eq. (4) as:

$$\frac{\mu_r}{\mu_\theta} r \frac{\partial}{\partial r}\left(r \frac{\partial}{\partial r}\right) E_z + \frac{\partial^2}{\partial \theta^2} E_z + r^2 k_0^2 \varepsilon_z \mu_r E_z = 0. \tag{5}$$

We can obtain the general solution of Eq. (5) by separation of variables:

$$E_z = E(r) e^{jn\theta}, n = 0, \pm 1, \pm 2, \ldots \tag{6}$$

$E(r)$ satisfies the following Bessel equation:

$$\frac{1}{E(r)} \frac{\mu_r}{\mu_\theta} r \frac{\partial}{\partial r}\left(r \frac{\partial}{\partial r}\right) E(r) - n^2 + r^2 k_0^2 \varepsilon_z \mu_r = 0. \tag{7}$$

Considering the medium of an ONM in this 2D concentric ring structure:

$$\mu_\theta = \frac{1}{\Delta}, \mu_r = \varepsilon_z = \Delta, \Delta \to 0. \tag{8}$$

We can rewrite Eq. (7) as:

$$\frac{1}{E(r)} \Delta^2 r \frac{\partial}{\partial r}\left(r \frac{\partial}{\partial r}\right) E(r) - n^2 + r^2 k_0^2 \Delta^2 = 0, \Delta \to 0. \tag{9}$$

In order to have a non-zero solution $E(r)$ in Eq. (9), it requires $n=0$, and hence Eqs. (6) and (9) can be reduced to

$$E_z = E(r), \tag{10}$$

$$\frac{1}{E(r)} r \frac{\partial}{\partial r}\left(r \frac{\partial}{\partial r}\right) E(r) + r^2 k_0^2 = 0. \tag{11}$$

Eq. (11) is a zero-order Bessel equation, and hence the solution to Eq. (4) in a concentric ring ONM can be written as:

$$E_z = A J_0(k_0 r) + B Y_0(k_0 r), \tag{12}$$

where $A$ and $B$ are constant, which can be fixed by the boundary condition that is related to the method of excitation. The eigenmode of this 2D concentric ring resonator is the linear superposition of the zero-order Bessel function and the zero-order Neumann function. That is the reason why we call it as the zero-order resonator. The resonance frequency is a continuous spectrum (no reflecting wall is utilized in this open resonator and the natural periodic boundary condition in the $\theta$ direction is removed by the ONM).

Next we will study the case when we set a unit line current in the center of this 2D concentric ring resonator. Then the electric field's $z$ component in each region (defined in Figure2(a)) can be written as:

$$\begin{cases} E_1(r,\omega) = A_1 J_0(k_0 r) + B_1 Y_0(k_0 r) \\ E_2(r,\omega) = A_2 J_0(k_0 r) + B_2 Y_0(k_0 r) \\ E_3(r,\omega) = A_3 H_0^{(1)}(k_0 r) + B_3 H_0^{(2)}(k_0 r) \end{cases}. \quad (13)$$

Note that the field in Region 2 is the eigenmode of the concentric ring ONM given in Eq. (12). The field in Region 1 contains a singularity as a line current source is set in the center ($B_1 \neq 0$). The field in Region 3 should be an outgoing cylindrical wave (no energy comes from beyond), and hence $B_3=0$ (the time harmonic is chosen as exp(-i$\omega$t)).

Considering the boundary condition that the electric field's tangential component should be continuous at the boundary, we can obtain:

$$\begin{cases} A_1 J_0(k_0 a) + B_1 Y_0(k_0 a) = A_2 J_0(k_0 a) + B_2 Y_0(k_0 a) \\ A_2 J_0(k_0 b) + B_2 Y_0(k_0 b) = A_3 H_0^{(1)}(k_0 b) \end{cases}. \quad (14)$$

where $a$ and $b$ are the inner and outer radii of the concentric ring, respectively (see Figure 2(a)). The magnetic field's tangential component can be determined from the electric field's $z$ component:

$$H_\theta = -\frac{1}{i\omega\mu_0\mu_\theta}\frac{\partial E_z}{\partial r}. \quad (15)$$

By combining Eqs. (8), (13) and (15), we can obtain ($B_3=0$):

$$\begin{cases} H_1(r,\omega) = -\dfrac{1}{i\omega\mu_0}\left[A_1 k_0 J_0'(k_0 r) + B_1 k_0 Y_0'(k_0 r)\right] \\ H_2(r,\omega) = -\dfrac{\Delta}{i\omega\mu_0}\left[A_2 k_0 J_0'(k_0 r) + B_2 k_0 Y_0'(k_0 r)\right] \overset{\Delta \to 0}{\sim} 0. \\ H_3(r,\omega) = -\dfrac{1}{i\omega\mu_0} A_3 k_0 H_0^{(1)\prime}(k_0 r) \end{cases} \quad (16)$$

Considering that the magnetic field's tangential component should also be continuous at the boundary, we can obtain:

$$\begin{cases} A_1 k_0 J_0'(k_0 a) + B_1 k_0 Y_0'(k_0 a) = 0 \\ A_3 k_0 H_0^{(1)\prime}(k_0 \rho) = 0 \end{cases}. \quad (17)$$

By combining Eqs. (14) and (17), we can obtain the coefficients in Eq. (13):

$$\begin{cases} A_2 = \dfrac{\left[J_0(k_0 a) - \dfrac{J_0'(k_0 a)}{Y_0'(k_0 a)} Y_0(k_0 a)\right]}{\left[J_0(k_0 a) - \dfrac{J_0(k_0 b)}{Y_0(k_0 b)} Y_0(k_0 a)\right]} A_1 \\ B_1 = -\dfrac{J_0'(k_0 a)}{Y_0'(k_0 a)} A_1 \\ B_2 = -\dfrac{J_0(k_0 b)}{Y_0(k_0 b)} \dfrac{\left[J_0(k_0 a) - \dfrac{J_0'(k_0 a)}{Y_0'(k_0 a)} Y_0(k_0 a)\right]}{\left[J_0(k_0 a) - \dfrac{J_0(k_0 b)}{Y_0(k_0 b)} Y_0(k_0 a)\right]} A_1 \\ A_3 = B_3 = 0 \end{cases} \quad (18)$$

One important feature from Eq. (18) is that the field in Region 3 is zero, which is consistent with the FEM simulation (see Figure 2(b)). Note that in numerical simulation, we use an extremely large number and an extremely small number (e.g. 1000 and 0.001) to approximately simulate the ONM (but not ideally infinity and zero). Thus, a very small electric field enters Region 3 in the simulation.

Note that to excite the eigenmode the line current is not necessarily at the center of the concentric ring: a deviated line current (see Figure 2 (c) and (d)) or a plane wave (see Figure 4(a)) can also excite the eigenmode of the resonator. The eigenmodes of the resonator are calculated with the FEM and shown in Figure 2(e) and (f), which corresponds to the zero-order Bessel function and the zero-order Neumann function in Eq. (12). The shape of the open resonator is not necessarily a concentric ring structure and can be designed in some other structures (e.g. in Figure 3). In practice, we can choose an open resonator of an appropriate shape to get high efficiency of excitation (e.g. a concentric ring cavity for a line current excitation, and a rectangular ring resonator for a plane wave excitation).

**Applications**

In addition to the traditional applications of an optical resonator (e.g. to confine electromagnetic energy), our open optical resonator designed with OST has many other novel applications. Here we list three important applications of the open resonator proposed in this paper. Firstly it can be utilized as an electromagnetic energy collector. As shown in Figure 2(d), we set five line currents in Region 1 outside the concentric ring resonator, and obtain a higher field inside the resonator (compared with a single line current case in Figure 2(b) and (c)).

Secondly, the resonator composed by the ONM can be of a subwavelength size (see Figure 5(a) and (b)), allowing it to function as an optical open micro-resonator for concentrating the light at the subwavelength scale.

Thirdly, we can achieve an electric field enhancement in a region of air after some small modifications of our open resonator. For example, we can cut off a small region from the concentric ring resonator in Figure 2(a) (i.e. the ring with a small air gap). As shown in Figure 6 (a), we can obtain an enhanced electric field in the air gap region of the ring resonator. Furthermore, we can acquire a higher field in the air gap region simply by adding more sources in the central air region of the ring resonator (see Figure 6(b)).

**Discussion and summary**

Optically open cavities/resonators composed of a complementary medium can be explained by TO: it can be treated as the folding of the space repeatedly along the $\theta$ direction. The effective optical path is zero (i.e. the space is folded to null), and there is no limitation on the resonance frequencies.

Some novel cavities/resonators have also been designed by TO [9-12]. All these cavities/resonators have a continuous resonance spectrum and do not need any reflecting walls. The reason for this can be easily understood: all these cavities/resonators are transformed from a free reference space to the real space (i.e. there is no cavity with reflecting walls in the reference space). This means that the resonance frequency is a continuous spectrum in the reference space. The coordinate transformation simply changes the field distribution but does not influence the resonance frequencies, and hence all these cavities/resonators have a continuous resonance spectrum. As an open resonator proposed in this paper, we connect two equivalent surfaces linked by the ONM together to form a closed loop filled with the ONM whose main axis is along the tangential direction of the two conformal boundaries of the loop. All the surfaces perpendicular to the main axis of the ONM inside the resonator are equivalent surfaces (e.g. surfaces perpendicular to the $\theta$ direction of a circular ring resonator), and hence when the light propagates along the tangential direction of the resonator (i.e. the direction of the ONM's main axis), there is no phase delay (i.e. the effective light path is zero). Note that The physical mechanism of the open resonator composed by ONM is similar to the open cavity composed by positive and negative refraction materials (i.e. the cancellation of the light path) [13, 14]. Actually a pair of positive and negative refraction materials perform equivalently like an ONM [18].

To further understand the physical mechanism of the open resonator composed by ONM, we make the following simulations of wave guiding (along the closed loop filled with the ONM) for 2D TE polarization case (i.e., the electric field is along the $z$ direction while the magnetic field is in the $x$-$y$ plane; consequently the light path should be determined by $n_r$ and $n_\theta$, besides the propagation length:

$$\begin{cases} n_r = \sqrt{\varepsilon_z \mu_\theta} \\ n_\theta = \sqrt{\varepsilon_z \mu_r} \end{cases}. \tag{19}$$

For an ONM, the effective refraction index is 1 and zero along $r$ and $\theta$ direction, respectively (e.g. $\mu_\theta=1/\Delta$, and $\mu_r = \varepsilon_z = \Delta$, where $\Delta \to 0$). If we replace $\mu_\theta$ by air while keeping other parameters unchanged (e.g. $\mu_r$ and $\varepsilon_z$ are both chosen as nearly zero), the effective refraction index of the ring resonator along $r$ direction decreases, and hence some energy inside the resonator will be leaked out into the surrounding air region (see Figure 7(b)). In this case, there is no resonance mode in a FDTD simulation (i.e. Q-factor drops to zero in Figure 7(b)).

If we replace $\mu_r$ by air, while keeping other parameters unchanged, the effective refraction index of the ring resonator along $\theta$ direction increases (see Eq. (19)). In this case, and the model pattern changes (see Figure 7(a) and (c)). Note that the resonance frequency and the Q-factor also slightly changes in the FDTD simulations. However we can still find a resonant mode in this case. In the FDTD simulation, the material dispersion is considered, and hence the resonant frequency is no longer continuous.

If we replace $\varepsilon_z$ by air, while keeping other parameters unchanged, both $n_r$ and $n_\theta$ increase. There is also some energy leakages from the resonator (see Figure 7(d)). In this case, the resonance mode also disappears in the FDTD simulation.

Now we can summarize different roles of $n_r$ and $n_\theta$ from above analyses: $n_r \rightarrow 1$ can prevent the energy leakage from the ring resonator or attract the wave from the surrounding space. $n_\theta \rightarrow 0$ can ensure the effective light path is zero along the $\theta$ direction and provide a high Q-value of the resonator.

In a practical application, the ONM can be approximately realized by the medium whose relative permittivity and permeability are far larger than 1 in the main axis direction (not necessary infinitely large) and between 0 and 1 (not necessarily zero) in other orthogonal directions, which corresponds to the case that $\Delta$ in Eq. (3) is not exactly zero. We also study the Q factor of the proposed resonator when $\Delta$ changes. As shown in Figure 8, the Q factor increases as $\Delta$ approaches zero (the Q is infinite if $\Delta=0$ for a lossless resonator in theory).

The zero-order optical resonator designed by an OST in this paper has some other special features. Firstly the method to design such a resonator is very simple: what we need to do is just design two conformal smooth closed curves according to the occasion of application (e.g. the method of the excitation), and fill an ONM inside this ring structure. Secondly, we only need one homogeneous medium (i.e. the ONM) to realize the resonator designed by the OST (e.g. without any gradient control), which is a significant advantage compared with other cavities/resonators design by TO. Thirdly, the resonator proposed here can be an optical open micro-resonator that confines the electromagnetic energy on a subwavelength scale. The proposed resonator can also perform as an electromagnetic energy collector or a power combination device (see Figure 2(d)). We can also achieve an electromagnetic energy squeezing effect in a region of air by cutting an air gap from the ring resonator (see Figure 6).

The novel optical open resonators proposed in this paper will have many applications (e.g. capturing electromagnetic waves, collecting electromagnetic energy, squeezing electromagnetic energy in a region of air , etc.) in the future.

**Acknowledgment**

This work is partially supported by the National High Technology Research and Development Program (863 Program) of China (No. 2012AA030402), the National Natural Science Foundation of China (Nos. 61178062 and 60990322), the Program of Zhejiang Leading Team of Science and Technology Innovation, Swedish VR grant (# 621-2011-4620) and AOARD.

**Author Contributions**

F. S. did calculations, and made simulations. F. S. and S. H. conceived the idea, wrote the article and revised it together. X. G. did FDTD simulations. All authors contributed to discussions. S. H. supervised this study and finalized the manuscript.

**Author Information**

Correspondence and requests for materials should be addressed to S. H. ( sailing@kth.se )

**Additional Information**

Competing financial interests: The authors declare no competing financial interests.

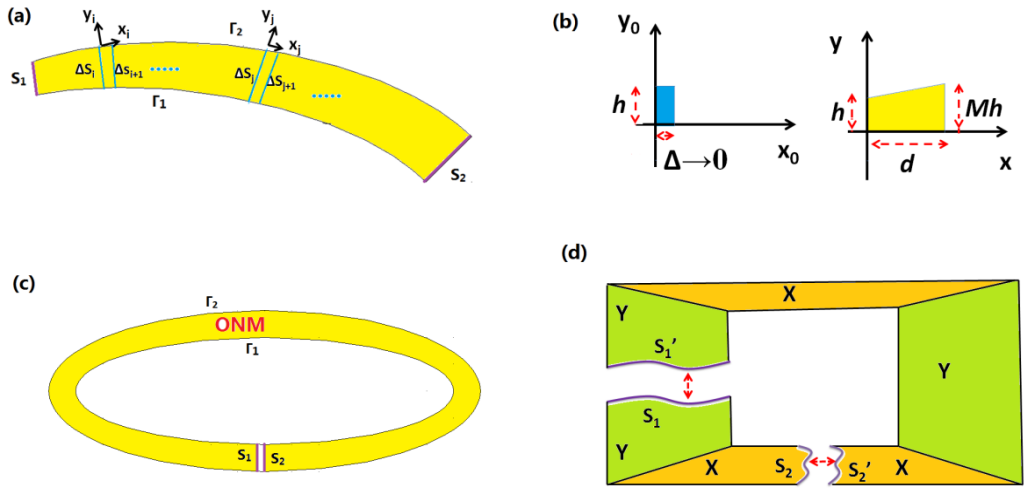

Figure 1| (a) Two arbitrarily shaped surfaces $S_1$ and $S_2$ as the terminated surfaces of two conformal smooth curves $\Gamma_1$ and $\Gamma_2$ (in accordance with the normal direction of these two smooth curves) will perform equivalently, if the ONM given in Eq. (2) is filled inside the each sub-region between $\Gamma_1$ and $\Gamma_2$ (the yellow region). (b) The coordinate transformation relation in each small sub-region. The left is the reference space $(x_0, y_0, z_0)$, and the right is the real space $(x, y, z)$ which corresponds to the local coordinate $(x_i, y_i, z_i)$ in (a). When $\Delta \to 0$, the blue volume in the reference space reduces to a surface which corresponds to the yellow region (i.e. the ONM) in the real space. If there is some scaling along the $y$ direction (i.e. $M \neq 0$), the main axis of the ONM is not along the $x$ direction (see Eq. (2)). If there is no compression and extension along $y$ direction (i.e. the normal direction of the conformal curves in (a)), we have $M=1$, which means the main axis of the ONM is along the $x$ direction (i.e. the tangential direction of the conformal curves in (a) and see Eq. (3)). (c) A closed loop is formed if we connect $S_1$ and $S_2$ together. (d) another way to design an optical resonator by the OST: $S_1$ and $S_2$ have been linked by the ONM. $S_1'$ and $S_2'$ have exactly the same shape as $S_1$ and $S_2$, respectively. $S_1'$ and $S_2'$ are also linked by the ONM. If we connect $S_1$ and $S_1'$, $S_2$ and $S_2'$ together, respectively, a closed loop filled with the ONM has been created. Such closed loop performs like an optical open resonator. The orange and green regions labeled by 'X' and 'Y' stand for the ONM with main axis along the $x$ and $y$ directions, respectively.

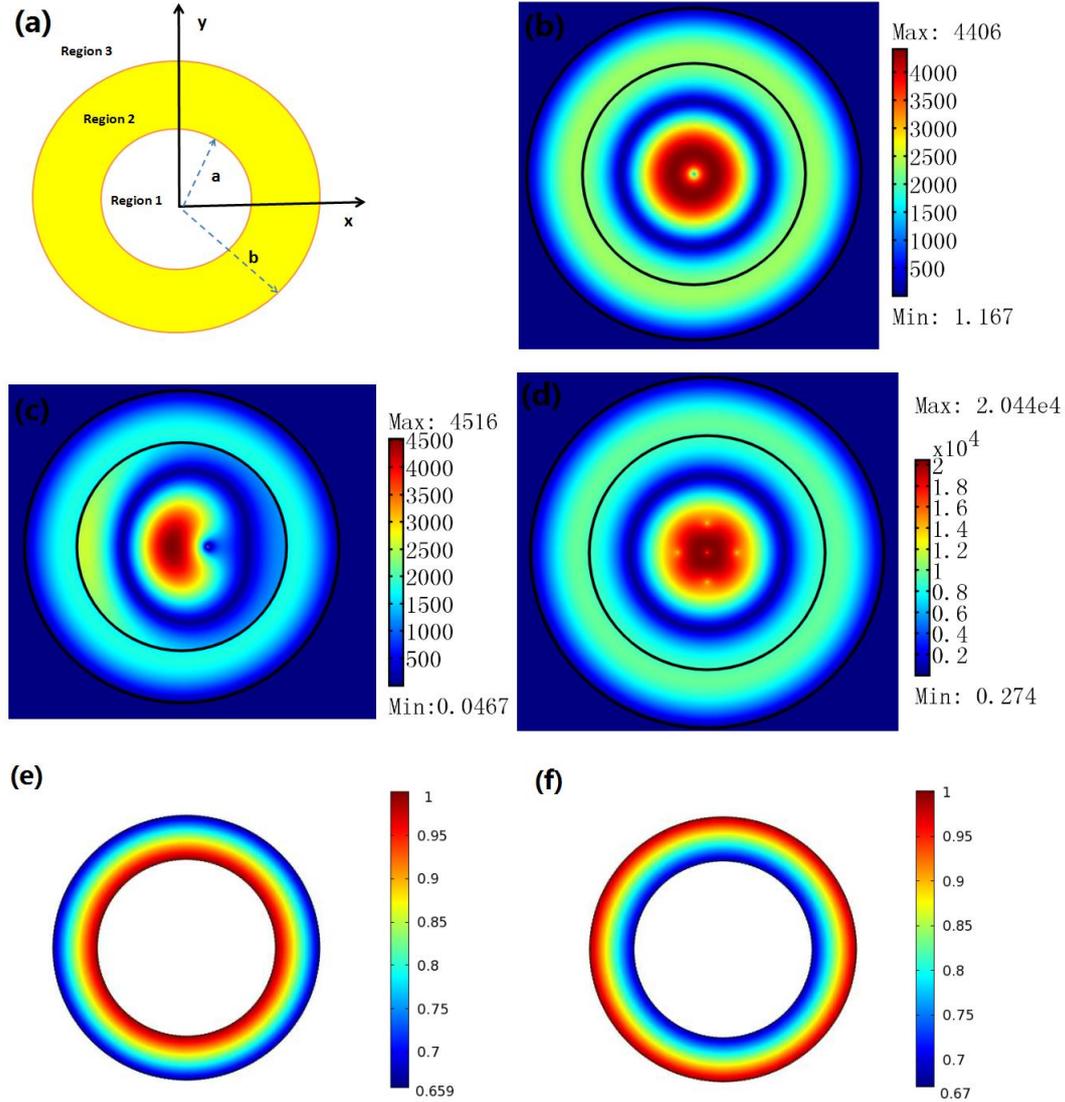

Figure 2| (a) A 2D open resonator structure we study here: regions 1 and 3 are free space. Region 2 is the concentric ring open resonator filled with the ONM whose main axis is in the $\theta$ direction. (b) to (d) are 2D FEM simulation results for a TE wave case. We plot the absolute value of the electric field's $z$ component. (b) and (c) we set a line current with amplitude 1A in the center and deviated $\lambda_0/6$ from the center of Region 1, respectively. (d) We set five line currents with amplitude 1A in Region 1. The size of the resonator is the same from (b) to (d): $a=2\lambda_0/3$, $b=\lambda_0$. (f) and (e) are two normalized eigenmodes of the resonator with the same size as (b) (obtained by the FEM numerical method), which correspond to the zero-order Bessel function and the zero-order Neumann function in Eq. (12), respectively.

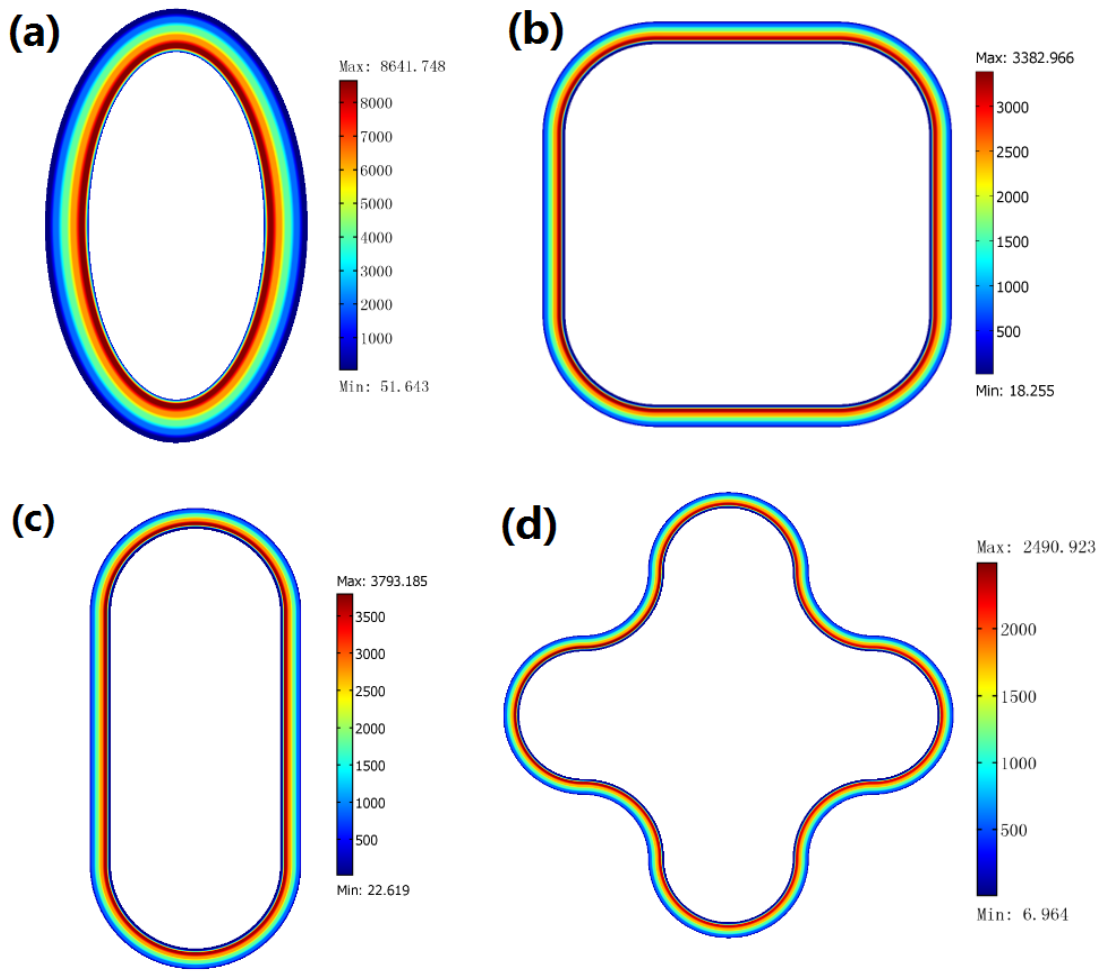

Figure 3| Some cavities of other shapes designed by the OST. We only plot the absolute value of electric field's *z* component inside the resonator for the TE polarization. We use one line current source with amplitude 1A inside the loop to excite the mode. (a) an elliptic resonator. (b) a rectangular resonator with smooth corners. (c) a playground shaped resonator. (d) a petaloid resonator. The whole size of these resonators are all smaller than the wavelength in above simulations.

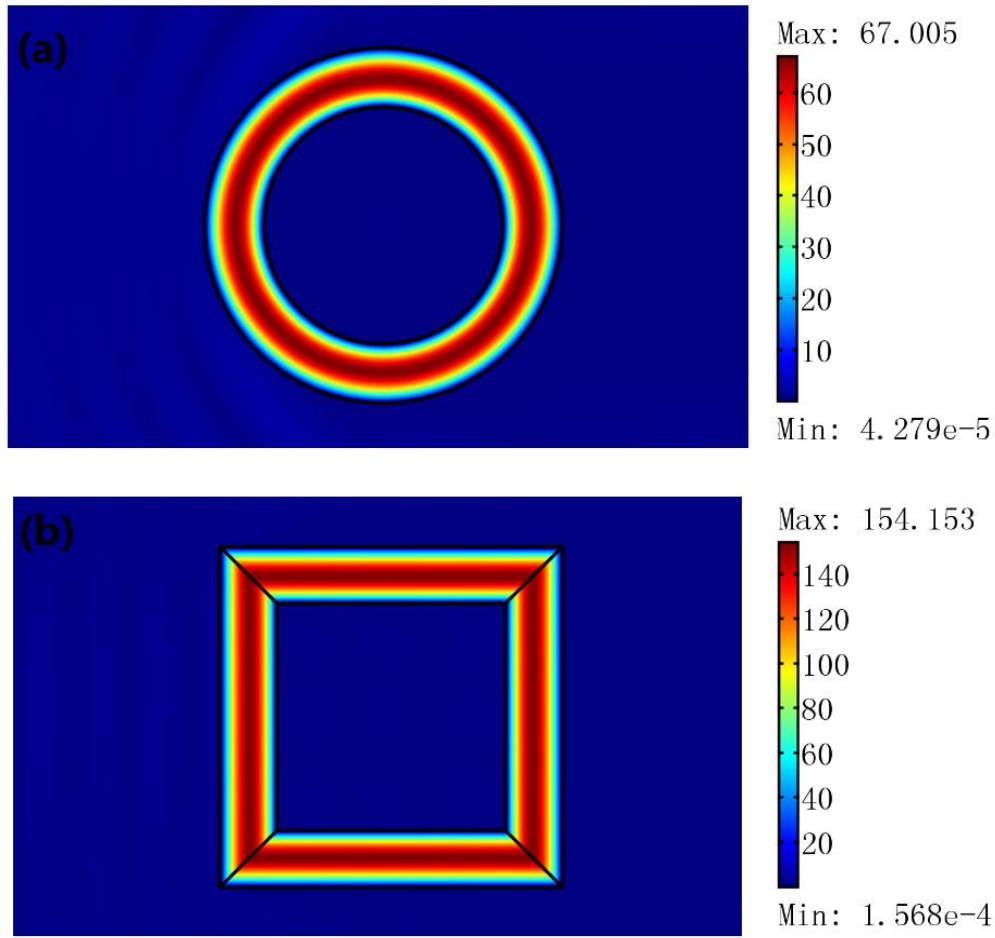

Figure 4| The 2D FEM simulation results for a TE wave case. A plane wave with unit amplitude is incident onto the resonator from the left. We plot the absolute value of the electric field's $z$ component. (a) A concentric ring resonator with inner radius $\lambda_0$ and outer radius $1.5\lambda_0$. (b) A rectangular ring resonator with inner side length of $2\lambda_0$ and outer side length of $3\lambda_0$.

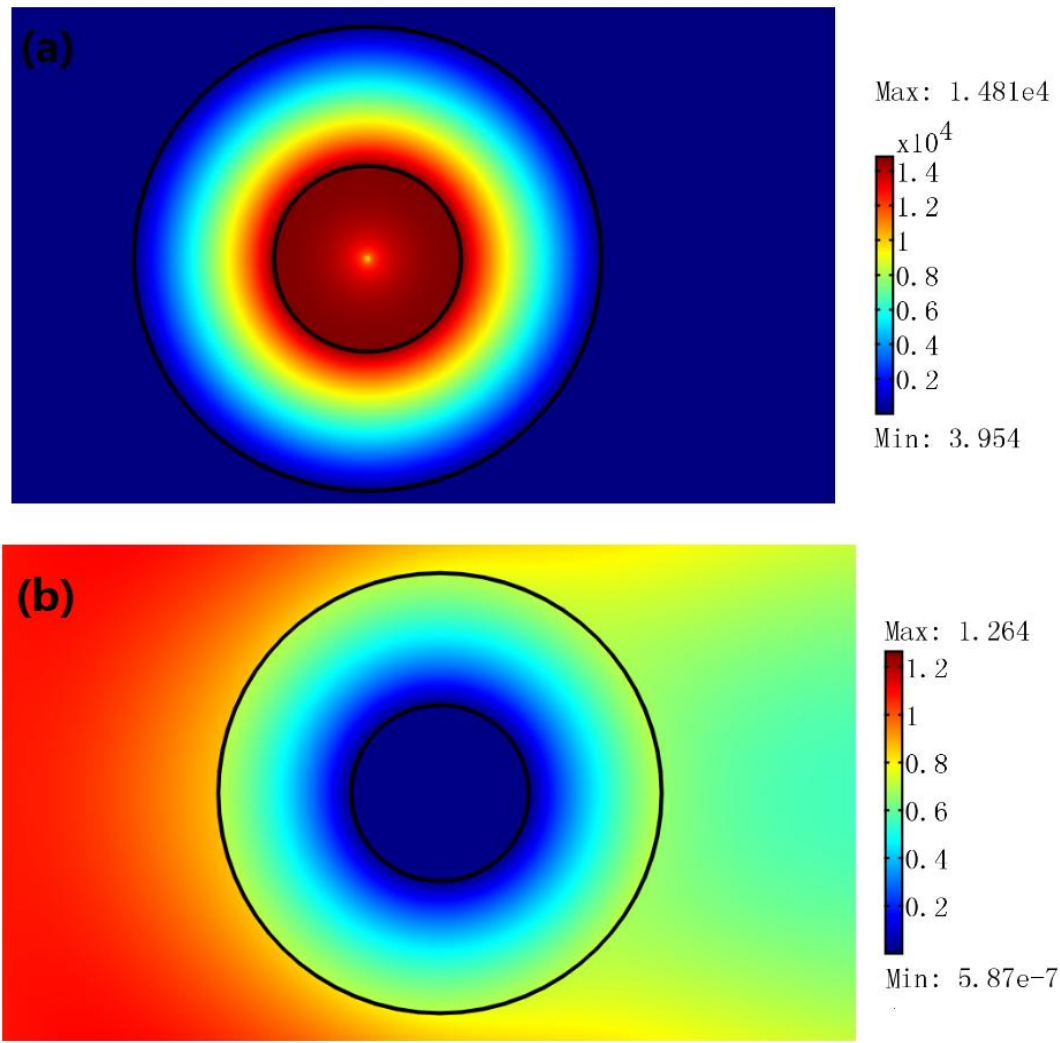

Figure 5| The 2D FEM simulation results for a TE wave case. We plot the absolute value of the electric field's $z$ component. The open concentric ring resonator here has a subwavelength size (e.g. $a=\lambda_0/15$, $b=\lambda_0/6$). (a) A line current with amplitude 1A is set at the center of the concentric ring resonator. (b) A plane wave with unit amplitude is incident onto the resonator from the left.

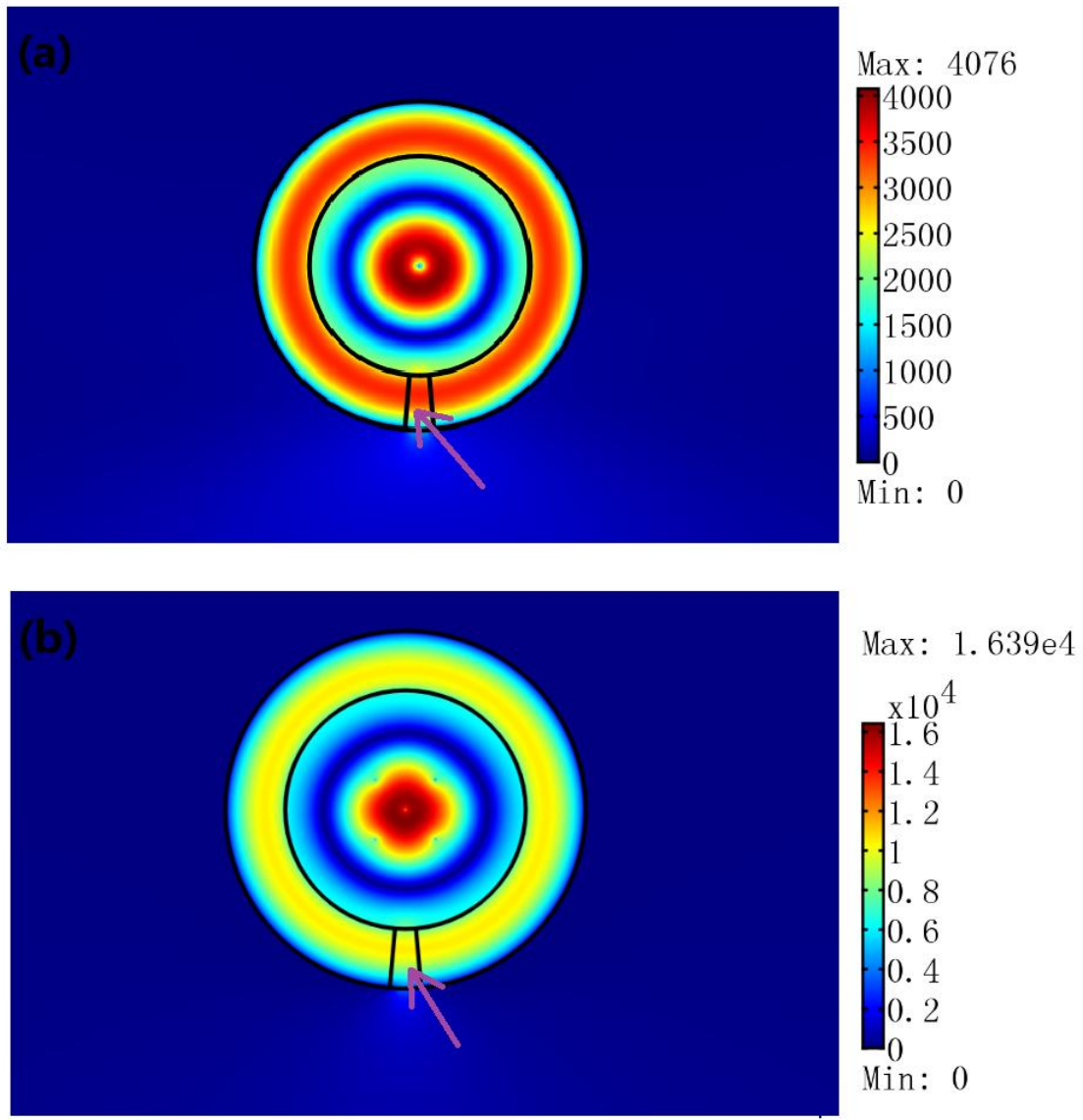

Figure 6| The 2D FEM simulation results for a TE wave case. We plot the absolute value of the electric field's *z* component. We cut off a small region from the concentric ring resonator in Figure 2(a). The inner and outer radii of the ring are $a=2\lambda_0/3$ and $b=\lambda_0$, respectively. (a) We set one line current with unit amplitude in the center of the ring. (b) We set five line currents with unit amplitude in the central air region of the ring resonator. The electric field is enhanced in the air gap region of the ring resonator. The pink arrow indicates the location of the air gap in the resonator.

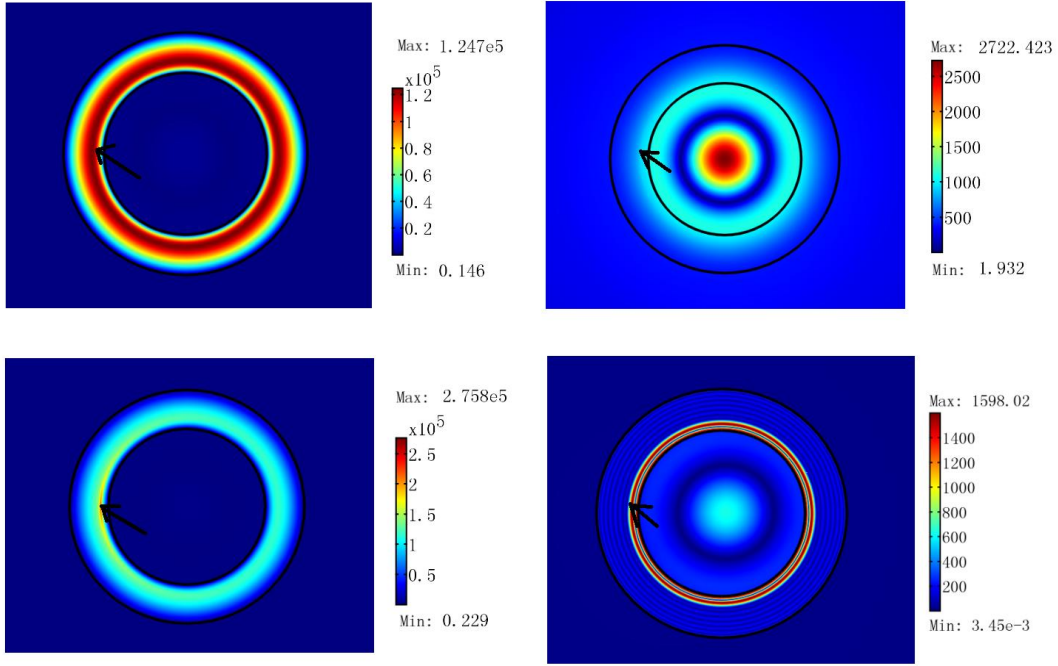

Figure 7| The 2D FEM simulation results for a TE wave case. We plot the absolute value of the electric field's $z$ component. We use one line current source with amplitude 1A inside the loop (indicated by the black arrow) to excite the mode. (a) The open circular resonator is composed by the ONM with $\mu_\theta = 10000$ and $\mu_r = \varepsilon_z = 0.0001$. (b) $\mu_\theta = 1$ and $\mu_r = \varepsilon_z = 0.0001$. (c) $\mu_\theta = 10000$, $\mu_r = 1$, and $\varepsilon_z = 0.0001$. (d) $\mu_\theta = 10000$, $\mu_r = 0.0001$, and $\varepsilon_z = 1$. The size of the cavity is chosen as $a = 2\lambda_0/3$, $b = \lambda_0$. We also use FDTD simulation to calculate the Q factor for each case. In the FDTD simulation, we assume the dispersion relation in the simulation to satisfy the required material parameters at the resonance frequency $\lambda_0$. Q factor in (a) and (c) are both very large around the resonant frequency. Q factors in (a) and (c) are very large (over million) at the resonant frequencies, while no resonance mode appears for (b) and (d) in the FDTD simulation.

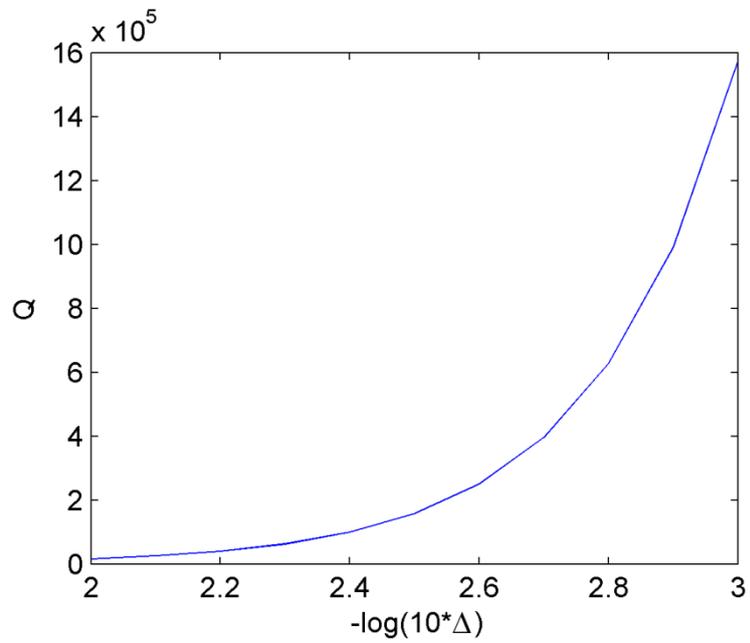

Figure 8| The FDTD simulation results: the relation between $\Delta$ and the Q factor of the resonator. Here we set $\mu_\theta=1/\Delta$, and $\mu_r =\varepsilon_z=\Delta$ in a concentric ring resonator with size $a=\lambda_0$ and $b=1.5\lambda_0$. $\lambda_0$ is the resonance wavelength. If $\Delta\rightarrow 0$ (e.g. $-\log(10*\Delta)$ approaches infinity), the Q factor approaches infinity.